# Visual Luminescence Thermometry Enabled by Phase-Transition-Activated Cross Relaxation of $Tb^{3+}$ Ions

L. Marciniak[1*], M. Szymczak[1*]

[1] *Institute of Low Temperature and Structure Research, Polish Academy of Sciences,*

*Okólna 2, 50-422 Wrocław, Poland*

\* corresponding author: l.marciniak@intibs.pl; m.szymczak@intibs.pl

**Abstract**

The development of visual luminescent thermometers capable of exhibiting pronounced color changes in response to temperature variations requires the rational design of phosphors with high spectrally selective thermal sensitivity. In this work, we present a strategy based on phase-transition-induced activation of cross-relaxation processes in $LiYO_2:Tb^{3+}$. The monoclinic-to-tetragonal structural phase transition modifies the point symmetry of $Tb^{3+}$ ions in the host lattice, enhances the Stark effect, and enables energetic resonance required for efficient cross relaxation. Consequently, emission originating from the $^5D_3$ excited state is rapidly quenched relative to that from the $^5D_4$ level above approximately 300 K, resulting in a distinct temperature-dependent color change of the emitted light from blue to green. This mechanism yields exceptionally high chromaticity-coordinate-based sensitivities, reaching $S_{Rx,max} = 0.40\%$ $K^{-1}$ and $S_{Ry,max}=0.72\%$ $K^{-1}$ at 410 K. Furthermore, phase-transition-driven modifications of the $Tb^{3+}$ emission spectral profile enable the realization of a multimode luminescent thermometer with a maximum relative sensitivity of $S_{Rmax}=13\%$ $K^{-1}$. The practical applicability of this system is demonstrated through an ON-OFF luminescent thermal switch and fully filter-free, dynamic two-dimensional thermal imaging using the blue and green channels of a standard digital camera, enabling intuitive visualization, remote readout, and temperature mapping under dynamic conditions.

**Introduction**

In materials undergoing a first-order structural phase transition, the thermal energy supplied during heating induces a change in crystal structure, typically accompanied by an increase in local symmetry and a concomitant change in entropy[1–8]. This transformation is associated with a reorganization of the local symmetry of the cationic sites. When these sites are occupied by luminescent centers, such as lanthanide ($Ln^{3+}$) or transition-metal (TM) ions, the structural transformation is reflected in pronounced changes in the luminescent properties of the material[9–12]. As demonstrated in recent studies, such effects can be effectively exploited for remote temperature sensing using luminescence thermometry[10,13–21]. Owing to the abrupt nature of the structural changes near the phase transition temperature exceptionally high relative sensitivities, reaching values as high as 30% K$^{-1}$, can be achieved [18]. The vast majority of luminescent thermometers based on structural phase transitions employ $Ln^{3+}$ ions as dopants, despite the fact that their spectroscopic properties are generally considered only weakly sensitive to changes in the local environment[22,23]. This limited sensitivity arises primarily from the shielding of the 4$f$ orbitals, responsible for the intraconfigurational electronic transitions governing $Ln^{3+}$ luminescence, by the outer 5$s$ and 5$p$ orbitals. Nevertheless, it has been shown that structural changes induced by a phase transition can manifest as variations in both luminescence kinetics and spectral characteristics. In particular, the phase transition modifies the Stark effect by altering both the number of Stark components into which the $Ln^{3+}$ energy levels are split and their energetic separation[12,21,24]. Consequently, the emission spectra exhibit distinct sets of Stark lines corresponding to the low-temperature (LT) and high-temperature (HT) phases of the host lattice.

As the temperature approaches the phase transition point, the emission lines associated with the LT phase undergo a rapid decrease in intensity, while those characteristic of the HT phase are simultaneously enhanced. This behavior is observed for all emission bands of $Ln^{3+}$ ions[18,20,25]. The temperature-induced spectral changes become increasingly pronounced with a decreasing number of Stark components and increasing crystal-field splitting strength, which leads to greater spectral separation between lines originating from the respective phases. Although the luminescence intensity ratio of two such emission lines can exhibit extremely high temperature sensitivity, accurate temperature readout is hindered by their close spectral proximity and narrow linewidths[10,17,26]. This limitation is particularly critical for thermal imaging applications, as it necessitates the use of band-pass filters with extremely narrow spectral bandwidths.

An alternative strategy involves exploiting thermally induced changes in the intensity ratio of two spectrally well-separated emission bands[15,21]. In phase transition-based luminescent thermometers, this approach is feasible when the luminescent ions exhibit hypersensitive transitions of a purely electric-dipole character, as is the case for $Eu^{3+}$ ions. In such systems, the emission band near 620 nm, corresponding to the $^5D_0 \rightarrow {}^7F_2$ electric-dipole transition, is highly sensitive to changes in the local crystal field, whereas the band near 590 nm, associated with the magnetic-dipole $^5D_0 \rightarrow {}^7F_1$ transition, remains largely insensitive to such a changes[27–30]. The spectral separation and linewidths of these bands permit the use of conventional band-pass filters for two-dimensional temperature imaging. However, the necessity of sequential filter switching limits the applicability of this method for monitoring dynamic temperature variations.

To overcome this limitation, a filter-free thermal imaging method based on the RGB channels of a digital camera has recently been proposed[15,21]. A comparison of the emission spectrum of $Eu^{3+}$-doped phosphors with the spectral response of the camera reveals that the red (R) channel records contributions from both the $^5D_0 \rightarrow {}^7F_1$ and $^5D_0 \rightarrow {}^7F_2$ transitions, whereas the green (G) channel predominantly captures the emission associated with the $^5D_0 \rightarrow {}^7F_1$ transition. This enables temperature mapping by acquiring a single image and calculating the ratio of the intensities recorded in the R and G channels. While this approach performs particularly well for phase transition-based thermometers, the partial contribution of the hypersensitive $^5D_0 \rightarrow {}^7F_2$ transition to both channels reduces the achievable temperature sensitivity. To enhance the relative sensitivity of filter-free thermal imaging, it is therefore necessary to develop phosphors exhibiting stronger and more selective temperature responses in distinct camera channels.

In this work, we propose a strategy that combines thermally activated cross-relaxation with phase transition-based luminescence thermometry in $LiYO_2:Tb^{3+}$. The $LiYO_2$ host undergoes a structural phase transition from a low-temperature monoclinic phase to a high-temperature tetragonal phase near room temperature (Figure 1a). Although this host material and the influence of the phase transition on the spectroscopic properties of $Tb^{3+}$ ions have been reported previously, earlier studies were limited in scope and focused on a single, relatively high dopant concentration, thereby suppressing emission from the $^5D_3$ level[18]. An increase in the point symmetry of the crystallographic site occupied by the $Tb^{3+}$ ions results in the reduction of the number of Stark components and the strength of the splitting of each multiplet (Figure 1b). This structural modification results in a distinct change in the emission behavior of $LiYO_2:Tb^{3+}$ across its phase transition. In the low-temperature (LT) phase, emissions from both the $^5D_3$ and $^5D_4$ excited states of $Tb^{3+}$ are observed. Due to the lack on energy matching the $\{^5D_3,{}^7F_6\}\leftrightarrow\{^5D_4,{}^7F_0\}$ cross-relaxation (CR) is effectively suppressed ("CR OFF"). However, in the high-temperature (HT) phase, the increased splitting of the $^7F_0$ state enhances the energy resonance conditions, thereby enabling CR ("CR ON") and leading to efficient non-radiative quenching of the $^5D_3$ level (Figure 1c). As a result, the onset of CR across the phase transition causes a significant reduction in the blue emission intensity relative to the green emission (Figure 1d). This abrupt spectral change occurring above the transition temperature induces a continuous thermal evolution in the shape of the LiYO$_2$:Tb$^{3+}$ emission spectrum (Figure 1e), accompanied by a perceptible shift in the emitted light color from blue to green (Figure 1f). As demonstrated in this study, this phenomenon enables the development of a visually discernible luminescent thermometer with high relative sensitivity (Figure 1g). Through systematic investigations, we demonstrate that the emission intensity recorded in the blue (B) channel of a digital camera, corresponding to emission from the $^5D_3$ level, can be effectively tuned relative to the green (G) channel emission originating from the $^5D_4$ level by controlling both the Tb$^{3+}$ concentration and the temperature (Figure 1h). The phase-transition-induced modifications of the Tb$^{3+}$ spectroscopic properties, together with the synergistic interaction between structural transformation and cross-relaxation, enable not only the realization of a multimode ratiometric luminescent thermometer but, more importantly, visual temperature sensing and imaging with the highest relative sensitivity reported to date based on chromaticity coordinates. This phenomenon is further exploited for dynamic, two-dimensional temperature mapping using a fully filter-free thermal imaging approach.

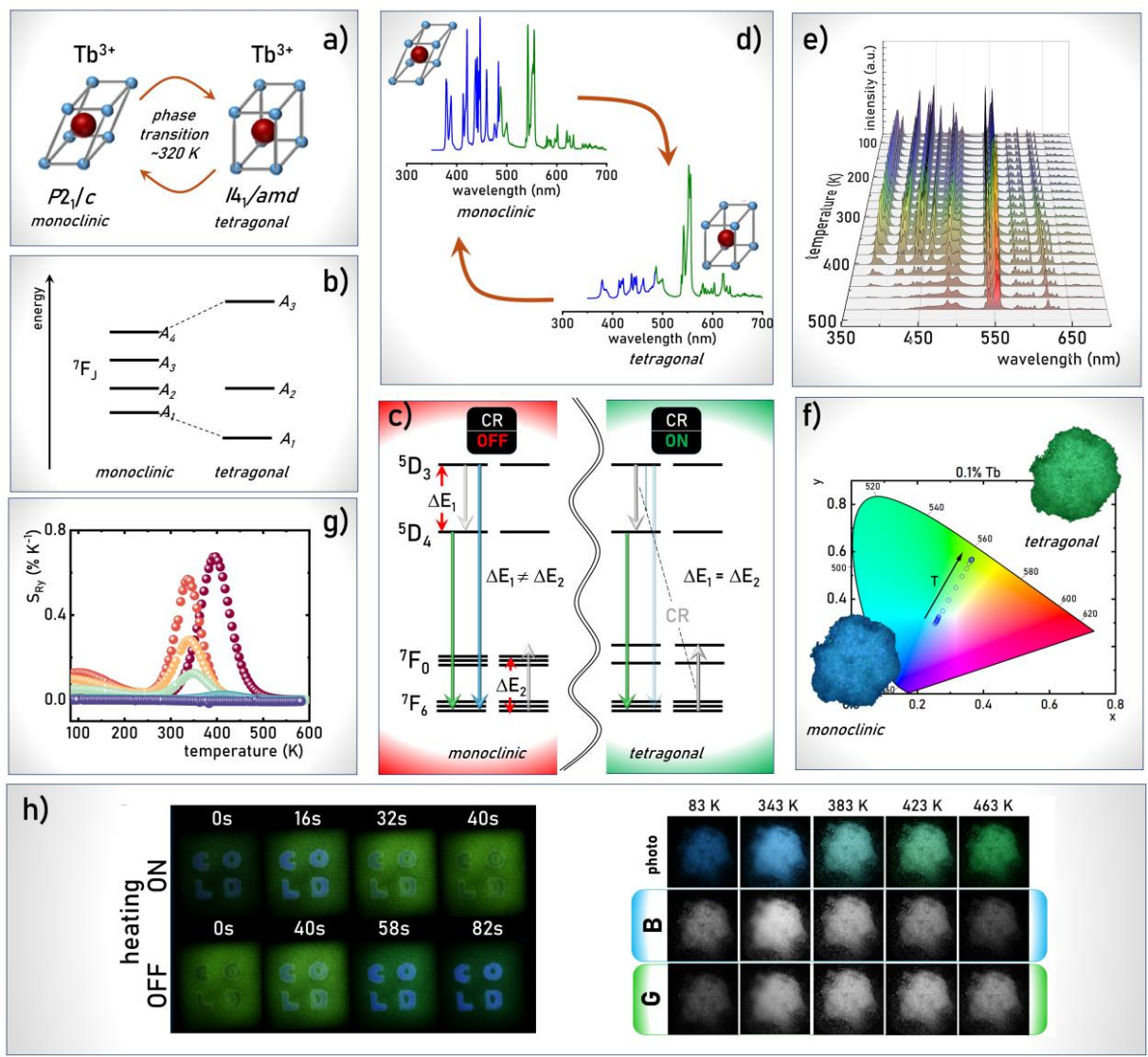

**Figure 1**. Schematic representation of the main concept of the work: the phase transition from low temperature monoclinic to high temperature tetragonal phases in LiYO$_2$ modifies the point symmetry occupied by the Tb$^{3+}$ ions from $P2_1/c$ to $I4_1/amd$ – a); an increase in the point symmetry results in the reduction of the number of Stark components into which given multiplet is split and the split strength – b); for LT phase of LiYO$_2$:Tb$^{3+}$ there is a lack of energy match for the $\{^5D_3,^7F_6\}\leftrightarrow\{^5D_4,^7F_0\}$ cross relaxation (CR) hence the CR is OFF resulting in the observation of both emission from $^5D_3$ and $^5D_4$ states, while for HT phase there is an energy match for CR (CR ON) and emission from $^5D_3$ is quenched – c); therefore the phase transition affects significantly the shape of the emission spectra of LiYO$_2$:Tb$^{3+}$ – d); the increase in temperature results in a progressive and efficient quenching of emission bands intensity from $^5D_3$ state in respect to those from $^5D_4$ – e); leading to the change of the emission colour from deep blue to the green – f); this change can be used for luminescence thermometry based on the chromatic coordinates of high sensitivity – g); and two dimensional dynamic thermal imaging – h).

**Experimental**

*Synthesis*

Powder samples of LiYO$_2$ doped with 0.1–5 mol% Tb$^{3+}$ were synthesized using a conventional high-temperature solid-state reaction route. Li$_2$CO$_3$ (99.9% purity, Chempur), Y$_2$O$_3$ (99.999% purity, Stanford Materials Corporation), and Tb$_4$O$_7$ (99.99% purity, Stanford Materials Corporation) were employed as starting materials. Stoichiometric amounts of the precursors were weighed and thoroughly mixed in an agate mortar with a few drops of hexane to ensure homogeneous blending. To compensate for lithium volatilization at elevated temperatures, a 3.5-fold excess of Li$_2$CO$_3$ was used. The resulting mixtures were transferred to alumina crucibles and annealed at 1273 K for 5 h under a reducing atmosphere, with a heating rate of 3 K min$^{-1}$. After natural cooling to room temperature, the obtained powders were reground to improve homogeneity and subsequently used for further structural and spectroscopic characterization.

*Characterization*

The obtained materials were examined by X-ray powder diffraction (XRD) using a PANalytical X'Pert Pro diffractometer in Bragg-Brentano geometry, using Ni-filtered Cu K$\alpha$ radiation (V = 40 kV, I = 30 mA). Measurements were made in the range of 10-90°, and the acquisition time was 30 min. Morphology of the samples were investigated using a scanning electron microscopy (SEM, FEI Nova NanoSEM 230). The samples were dispersed in alcohol, and a drop of the suspension was placed onto a carbon stub. SEM images were then collected using an accelerating voltage of 5.0 kV.

Raman spectra were collected using an Edinburgh Instruments RMS-1000 Raman microscope equipped with a 532 nm excitation laser, a 20x microscope objective, and an 1800 lines mm$^{-1}$ diffraction grating. Temperature control during the measurements was ensured using a THMS600 heating-cooling stage from Linkam.

Excitation and emission spectra were measured using an FLS1000 spectrometer from Edinburgh Instruments, equipped with a 450 W xenon lamp as the excitation source, and an R928P side window photomultiplier tube from Hamamatsu as a detector. The luminescence decay curves were recorded on this spectrometer with a pulsed excitation, provided by a μFlash lamp. Temperature of the samples during the measurements was also controlled using THMS600 heating-cooling stage from Linkam.

The luminescence decay profiles were fitted with a double exponential function, as expressed below:

$$I(t) = I_0 + A\exp\left(-\frac{t}{\tau_1}\right) + A\exp\left(-\frac{t}{\tau_2}\right) \qquad (1)$$

where $I(t)$, $A_1$, $A_2$, $y_0$, $\tau_1$, and $\tau_2$, represent the luminescence intensity at time $t$ right after laser pulse excitation, preexponential factors, offset, and decay parameters, respectively. Based on these parameters the average lifetime $\tau_{avr}$ were calculated as follows:

$$\tau_{avr} = \frac{A_1\tau_1^2 + A_2\tau_2^2}{A_1\tau_1 + A_2\tau_2} \qquad (2)$$

Photographs documenting the luminescence of the samples, as well as all images acquired for both proof-of-concept experiments, were collected using a Canon EOS 400D camera equipped with an EF-S 60 mm macro lens and a long-pass optical filter. Excitation was provided by a xenon lamp integrated in an Edinburgh Instruments FLS1000 spectrofluorometer; the excitation wavelength was set to 285 nm.

For both proof-of-concept experiments, the phosphor powders were dispersed in a high-viscosity silicone paste (KORASILON-Paste, Kurt Obermeuer GmbH &Co. KG) to obtain mechanically stable coatings. In experiment 1, time-lapse photographs were acquired during heating and subsequent cooling of the metal element of a CELMA OP2000P heat gun (rated power 2000 W). Heating was performed with the set temperature of 450 °C and an air flow rate of 250 L min$^{-1}$, whereas cooling was carried out using a 50 °C mode with an air flow rate of 500 L min$^{-1}$. The camera exposure time was 8 s for each photo. In experiment 2 (RGB-based mapping), images were collected using an exposure time of 15 s. The temperature was monitored using a FLIR T540 thermographic camera with a measurement accuracy of ±0.5 °C.

The red, green, and blue (RGB) channels were extracted from the acquired photographs using IrfanView (v4.51, 64-bit). The resulting channel-specific intensity maps were further processed in ImageJ (v1.8.0_172), where pixel-by-pixel ratio images were calculated by dividing selected channels to generate ratiometric maps.

**Results and discussion**

Temperature constitutes one of the most fundamental thermodynamic parameters capable of driving reversible structural phase transitions in crystalline probes. LiYO$_2$ represents a prototypical example of such behavior, crystallizing in two distinct polymorphs depending on experienced temperature: a low-temperature (LT) monoclinic phase (space group *P2$_1$/c*) and a high-temperature (HT) phase adopting a higher-symmetry tetragonal structure (space group *I4$_1$/amd*) (Figure 2a)[25,31–36]. The temperature-induced transformation between these polymorphs is accompanied by profound structural rearrangements, encompassing modifications of interatomic bond lengths, coordination geometries, and the long-range connectivity of polyhedral building units within the crystal lattice. In both structures, Y$^{3+}$ ions are coordinated by six oxygen atoms, forming YO$_6$ octahedra; however, the extent of octahedral distortion differs markedly between the two phases. In the LT monoclinic structure, the YO$_6$ units exhibit pronounced distortion, with all Y$^{3+}$-O$^{2-}$ bond lengths being inequivalent and spanning a broad range from 2.208 to 2.341 Å[31,32]. This large dispersion reflects a low-symmetry local environment around the Y$^{3+}$ ions. Upon transformation to the HT tetragonal phase, the YO$_6$ octahedra undergo partial regularization: four Y$^{3+}$-O$^{2-}$ bonds converge to a length of 2.241 Å, while the remaining two bonds are elongated to 2.313 Å. The resulting reduction in bond-length dispersion directly contributes to the enhancement of crystallographic symmetry at elevated temperatures. Simultaneously, a subtle contraction of the average Y$^{3+}$-O$^{2-}$ bond length is observed, decreasing from 2.2765 Å in the LT phase to 2.265 Å in the HT phase. Beyond the immediate coordination sphere, substantial differences between the two polymorphs are manifested in the extended coordination environment of Y$^{3+}$ ions, which involves lithium-oxygen polyhedra linked to the YO$_6$ octahedra via shared oxygen atoms. In the LT phase, each Y$^{3+}$ ion is connected through oxygen bridges to seven Li$^+$ ions, each characterized by threefold oxygen coordination. In contrast, the HT structure exhibits increased connectivity, with Y$^{3+}$ ions linked to eight Li$^+$ ions, each adopting a fourfold oxygen coordination. This pronounced reorganization of the lithium sublattice plays an important role in stabilizing the high-symmetry tetragonal framework at elevated temperatures and significantly alters the lattice rigidity and vibrational landscape.

Such temperature-driven structural rearrangements are of particular relevance for luminescent materials, as the optical response of activator ions is highly sensitive to variations in the local crystal field. In the present study, Tb$^{3+}$ ions were employed as luminescent probes through substitution at Y$^{3+}$ sites, a process facilitated by their identical charge state and closely matched ionic radii. X-ray diffraction analysis of LiYO$_2$ samples doped with 0.1-5 mol% Tb$^{3+}$ confirms the formation of the monoclinic phase and preservation of the host crystal structure across the entire compositional range, with no evidence of secondary phase formation induced by doping (Figure 2b). Complementary scanning electron microscopy reveals the formation of microcrystalline particles with irregular morphologies, consistent with previously reported characteristics of LiYO$_2$-based materials synthesized by solid-state reaction (Figure 2c).

To elucidate the influence of the structural phase transition on local bonding interactions and lattice dynamics, temperature-dependent Raman spectroscopy was conducted in the range of 298-373 K with 5 K increments for all the samples. Representative Raman spectra acquired for the LiYO$_2$:0.1% Tb$^{3+}$ sample exhibit a pronounced simplification upon heating, manifested by a marked reduction in the number of Raman-active modes (Figure 2d). This spectral evolution becomes evident above approximately 318 K and unambiguously signals the onset of a higher-symmetry phase, in full agreement with literature reports and consistent with the formation of the tetragonal HT polymorph[37]. At temperatures exceeding ~318 K, a progressive attenuation of the Raman bands located near 140 cm$^{-1}$ and 480 cm$^{-1}$ is observed. This behavior corroborates the structural regularization accompanying the monoclinic-to-tetragonal phase transition. In particular, the Raman features centered around 480 and 520 cm$^{-1}$ are assigned to internal vibrations of the YO$_6$ octahedra, corresponding to O$^{2-}$-Y$^{3+}$-O$^{2-}$ bending and Y$^{3+}$-O$^{2-}$ stretching modes, respectively. Their temperature-dependent evolution reflects the decreasing distortion of the YO$_6$ coordination units and the concomitant narrowing of the Y$^{3+}$-O$^{2-}$ bond-length distribution in the HT phase. The low-frequency band at approximately 140 cm$^{-1}$ is probably attributed to external lattice modes, involving either translational motions of Y$^{3+}$ ions or rotational and liberational vibrations of the YO$_6$ octahedra (Figure 2e). These modes are strongly coupled to Li$^+$ ions through Y$^{3+}$-O$^{2-}$-Li$^+$ bonds, whose connectivity and coordination undergo significant reconfiguration during the monoclinic to tetragonal phase transition. In the HT phase, the increased number of Li$^+$ ions linked to Y$^{3+}$ via shared oxygen atoms, together with the higher oxygen coordination of Li$^+$, leads to a pronounced stiffening of the crystal lattice. Consequently, the amplitude of low-frequency lattice vibrations is suppressed, giving rise to the observed weakening of the corresponding Raman modes. Analysis of the temperature-dependent Raman spectra for all investigated samples with various Tb$^{3+}$ doping enabled the determination of the phase transition temperature T$_{PT}$ (Figure 2f). Notably, increasing the $Tb^{3+}$ concentration results in a shift of $T_{PT}$ toward higher temperatures. This effect can be rationalized by the larger ionic radius of $Tb^{3+}$ relative to $Y^{3+}$, which introduces local lattice expansion, thereby stabilizing the low-temperature monoclinic phase[16,20,21]. As a result, a higher thermal energy input is required to overcome this stabilization and induce the transition to the tetragonal HT phase.

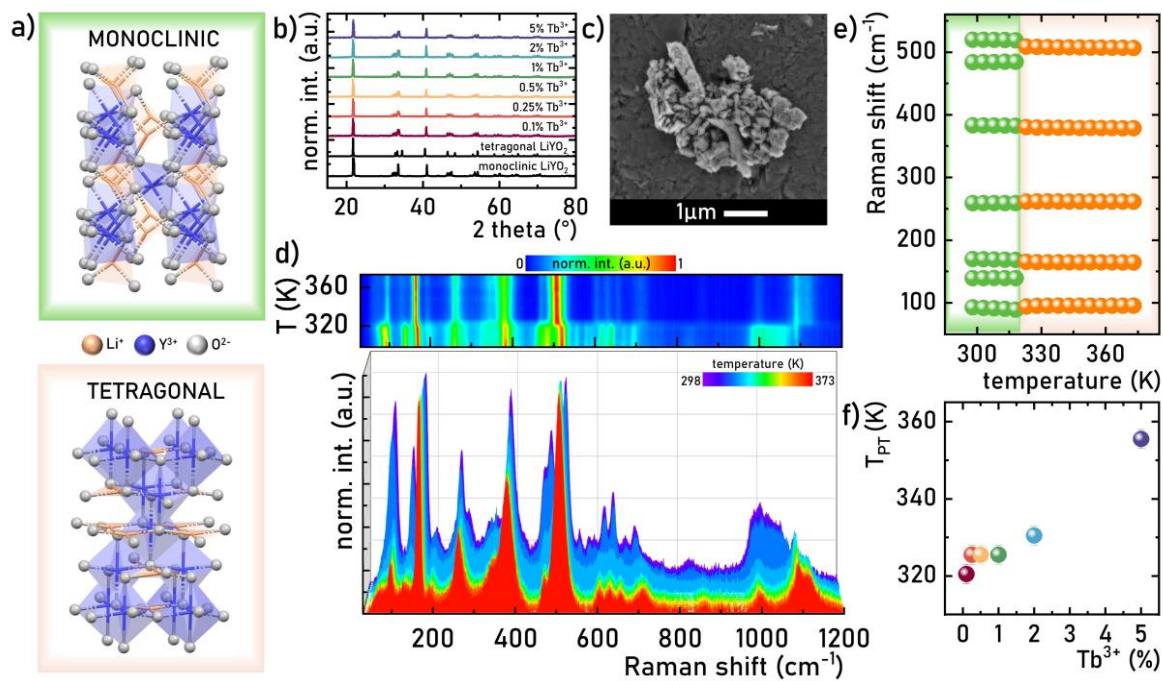

**Figure 2**. Monoclinic and tetragonal structures of $LiYO_2$ – a); comaprisson of reeom temperature XRD patterns of $LiYO_2:Tb^{3+}$ with different $Tb^{3+}$ ions concentration – b); representative SEM image for $LiYO_2:0.1\%Tb^{3+}$ – c); thermal map of normalized Raman spectra of $LiYO_2:1\%Tb^{3+}$ – d); thermal evolution of energies of modes of $LiYO_2:1\%Tb^{3+}$ – e); the influence of $Tb^{3+}$ ions concentration on the phase transition temperature ($T_{PT}$) of $LiYO_2:Tb^{3+}$ – f);

Phosphors doped with $Tb^{3+}$ ions are well known for their intense luminescence arising from the radiative depopulation of the $^5D_4$ excited state to the $^7F_J$ manifold (Figure 3a)[38,39]. The highest emission intensities are observed for bands centered at approximately 490 nm and 545 nm, corresponding to the $^5D_4 \rightarrow {^7F_6}$ and $^5D_4 \rightarrow {^7F_5}$ electronic transitions, respectively.

Additional, significantly weaker emission bands appear near 590 nm ($^5D_4 \rightarrow {}^7F_4$) and 630 nm ($^5D_4 \rightarrow {}^7F_3$). In host materials characterized by low phonon energies and low $Tb^{3+}$ concentrations, emission originating from the higher-lying $^5D_3$ level can also be observed in the UV-blue spectral range. The relatively rare occurrence of $^5D_3$ emission is primarily related to the relatively small energy gap between the $^5D_3$ and $^5D_4$ levels, approximately 6000 cm$^{-1}$. In high-phonon-energy host lattices, such as phosphates or borates, this energy gap can be efficiently bridged by about four host phonons, enabling rapid non-radiative relaxation[40–43]. Moreover, this energy separation closely matches the gap between the $^7F_0$ and $^7F_6$ levels, which strongly promotes depopulation of the $^5D_3$ level via cross-relaxation processes of the type $\{^5D_3, {}^7F_0\} \leftrightarrow \{^5D_4, {}^7F_6\}$. The probability of such cross-relaxation increases with shortening of the average distance between interacting $Tb^{3+}$ ions, leading to a rapid decrease in the intensity of the $^5D_3$-related emission bands with increasing dopant concentration. A comparison of the emission spectra of $LiYO_2$:0.1%$Tb^{3+}$ measured at 83 K and 360 K, corresponding to the low-temperature (LT) and high-temperature (HT) structural phases of $LiYO_2$:$Tb^{3+}$, respectively, reveals pronounced differences arising from both thermal effects and the structural phase transition of the host lattice (Figure 3b). Most notably, the relative intensity of the emission bands associated with the depopulation of the $^5D_3$ level decreases with respect to those originating from the $^5D_4$ level in the HT phase. As discussed below, this behavior is attributed to thermally activated cross-relaxation processes facilitated by the phase transition of the host material. In addition to the expected thermal broadening of the Stark components constituting each emission band, a reduction in the number of observable Stark lines is evident in the HT phase. Simultaneously, an increased energetic separation of these lines is observed, indicating enhanced crystal-field splitting. This behavior reflects the higher local symmetry of the HT phase, which is commonly associated with an increased crystal-field strength. The temperature-induced shifts in the spectral positions of the Stark components are particularly evident in the normalized luminescence maps (Figure 3c).

With increasing $Tb^{3+}$ concentration, a systematic reduction in the contribution of emission from the $^5D_3$ level relative to that from the $^5D_4$ level is observed, consistent with the growing efficiency of the $\{^5D_3, ^7F_0\} \rightarrow \{^5D_4, ^7F_6\}$ cross-relaxation process (Figure 3d). While, for $LiYO_2$:0.1%$Tb^{3+}$, the emission intensities associated with the $^5D_3 \rightarrow ^7F_J$ and $^5D_4 \rightarrow ^7F_J$ transitions are comparable at 83 K, the $^5D_3$-related emission becomes barely detectable at 2%$Tb^{3+}$. To quantitatively evaluate this effect, the luminescence intensity ratio $LIR_1$ was determined a follows:

$$LIR_1 = \frac{\int_{350nm}^{510nm} \left(^5D_3 \rightarrow ^7F_J\right) d\lambda}{\int_{530nm}^{700nm} \left(^5D_4 \rightarrow ^7F_J\right) d\lambda} \qquad (3)$$

For a 0.1%$Tb^{3+}$ the emission intensity of the $^5D_3 \rightarrow ^7F_J$ transitions slightly exceeds that of the $^5D_4 \rightarrow ^7F_J$ transitions, yielding an $LIR_1$ value of 1.4 (Figure 3e). Increasing the $Tb^{3+}$ concentration to 0.5% $Tb^{3+}$ results in a twofold reduction of $LIR_1$, while further increases in dopant concentration lead to a rapid decrease of this parameter. Analysis of the luminescence decay kinetics of both emitting levels, performed at 83 K, further corroborates the role of cross-relaxation (Figure 3f). The average lifetime ($\tau_{avr}$) of the $^5D_3$ level shortens markedly with increasing $Tb^{3+}$ concentration, from approximately 0.9 ms for 0.1%$Tb^{3+}$ to 0.3 ms for 5%$Tb^{3+}$. In contrast, the decay kinetics of the $^5D_4$ level remain largely unaffected by changes in $Tb^{3+}$ concentration, with $\tau_{avr} \approx 1.5$ ms across the investigated dopant concentration range. The remarkable stability of the $\tau_{avr}$ of the $^5D_4$ arises from two key factors. First, there is no energetically match for cross-relaxation pathway leading to depopulation of this level. Second, the large energy separation between the $^5D_4$ level and the underlying $^7F_0$ level (>12,000 cm$^{-1}$) effectively suppresses nonradiative multiphonon relaxation[40,43]. Consequently, the absence of additional nonradiative depopulation channels for the $^5D_4$ level at higher $Tb^{3+}$ concentrations leads to a redistribution of the excited-state population between the $^5D_3$ and $^5D_4$ levels, resulting in a progressive enhancement of the luminescence intensity associated with the $^5D_4$ state.

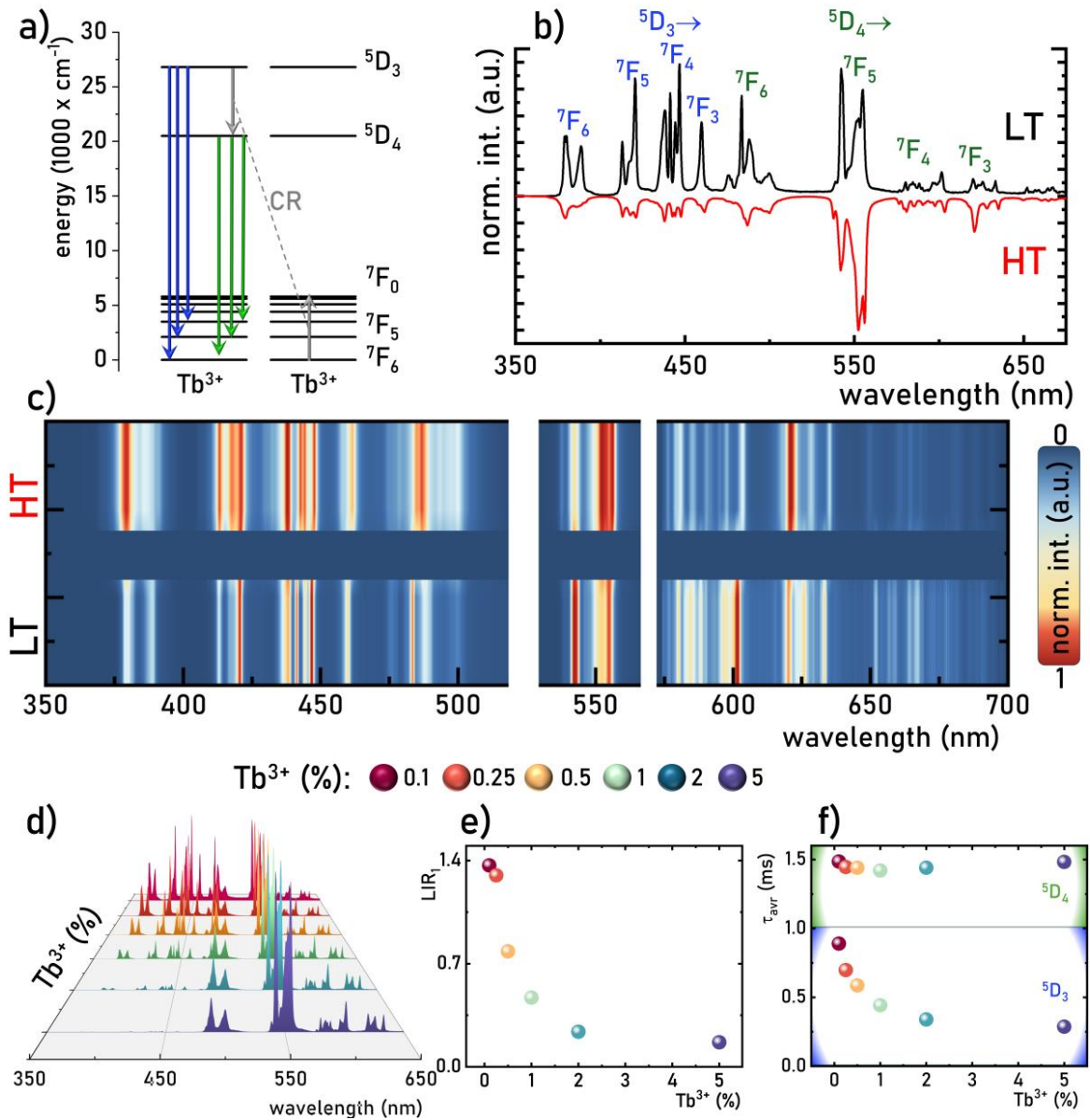

**Figure 3**. Simplified energy diagram of $Tb^{3+}$ ions – a); normalized emission spectra of $LiYO_2$:0.1%$Tb^{3+}$ representative for low temperature (measured at 83 K) and high temperature (at 363 K) – b); maps of emission spectra for LT and HT phases of $LiYO_2$:0.1%$Tb^{3+}$ normalized in three different spectral ranges – c); the normalized emission spectra of $LiYO_2$:$Tb^{3+}$ for different concentrations of $Tb^{3+}$ upon $\lambda_{exc}$ = 285 nm – d), the influence of $Tb^{3+}$ ions concentration of $LIR_1$ – e) and $\tau_{avr}$ of $^5D_3$ and $^5D_4$ states – f) measured at 83 K.

To investigate the influence of temperature on the luminescent properties of LiYO$_2$:Tb$^{3+}$, emission spectra were recorded over a broad temperature range from 83 to 703 K. The emission spectra obtained for LiYO$_2$:0.1%Tb$^{3+}$ (Figure 4a, Figure S1, S2) reveal that, initially at low temperatures, increasing temperature does not lead to a significant reduction in the overall luminescence intensity of Tb$^{3+}$ ions. However, above approximately 323 K, a pronounced quenching in the emission intensity originating from the $^5D_3$ level is observed relative to the emission from the $^5D_4$ level. A detailed analysis of the temperature dependence of the luminescence intensity associated with the $^5D_3 \rightarrow {^7F_J}$ transitions, normalized to the total emission intensity, indicates that heating up to approximately 320 K has a negligible effect on the population of the $^5D_3$ level (Figure 4b). Above this temperature, a further increase in temperature initially results in a modest enhancement of the $^5D_3$ emission intensity by approximately 10%, followed by a rapid quenching of its intensity with continued heating. This behavior is also observed for higher Tb$^{3+}$ concentrations; however, due to concentration-dependent cross-relaxation, the initial contribution of the $^5D_3$-related emission to the total luminescence spectrum decreases for higher dopant concentration. The pronounced thermal quenching of the $^5D_3$ emission observed above ~320 K suggests that the cross-relaxation process responsible for depopulation of this state is thermally activated. Such behavior typically occurs when the energetic resonance required for cross-relaxation is achieved through thermal population of Stark sublevels. Nevertheless, the strong correlation between the onset temperature of rapid $^5D_3$ quenching and the structural phase transition temperature of the LiYO$_2$ host indicates that a different mechanism plays a dominant role. Specifically, the phase transition to the HT phase, accompanied by an increase in local symmetry, leads to stronger splitting of the $^7F_J$ multiplets into Stark components. This enhanced crystal-field splitting facilitates the energetic resonance necessary for the $\{^5D_3, {^7F_0}\} \rightarrow \{^5D_4, {^7F_6}\}$ cross-relaxation process. In addition, the structural transformation from the monoclinic to the tetragonal phase of the LiYO$_2$ shortens the average Y$^{3+}$-Y$^{3+}$ distance from 3.4647 Å to 3.4157 Å, further increasing the probability of cross-relaxation. The increase in luminescence intensity observed above approximately 300 K can be attributed to phase-transition-induced modifications of the excitation spectrum of LiYO$_2$:Tb$^{3+}$ (Figure 4c, Figure S3 see also Figure S4 and S5). In the ultraviolet region, the excitation spectrum of Tb$^{3+}$ ions is dominated by $4f^8 \rightarrow 4f^75d^1$ transitions, the spectral shape and position of which are highly sensitive to changes in the local coordination symmetry. The structural phase transition results in a redshift of the excitation band, leading to an increased absorption cross-section at the excitation wavelength and, consequently, enhanced luminescence intensity. This effect is observed for emission originating from both the $^5D_3$ and $^5D_4$ levels (Figure 4d). The thermal evolution of the luminescence intensities associated with these two emitting levels can be divided into three distinct temperature regions (I-III in Figure 4d). In the low-temperature range (I) from approximately 83 to 300 K, only minor thermal variations are observed, although the $^5D_3$ emission exhibits slightly stronger thermal quenching than the $^5D_4$ emission. In the intermediate temperature range (II) above 300 K, both emission intensities increase as a result of the enhanced absorption cross-section induced by the phase transition. For the $^5D_4$ level, the luminescence intensity increases monotonically up to approximately 400 K, reaching about 180% of its initial value at 83 K. In contrast, the emission from the $^5D_3$ level increases only up to approximately 350 K, beyond which further heating triggers rapid thermal quenching driven by cross-relaxation. In the case of the emission from $^5D_4$ level, a decrease in intensity is observed only at temperatures exceeding 400 K (III). Analysis of the temperature dependence of the luminescence decay kinetics further supports these conclusions. The $\tau_{avr}$ of the $^5D_4$ level remains remarkably stable up to approximately 600 K, confirming that the observed intensity variations originate from changes in excitation and population efficiency rather than from thermally activated nonradiative depopulation pathways (Figure S6 and S7). In contrast, the pronounced shortening of the $\tau_{avr}$ associated with the $^5D_3$ level above 300 K provides direct evidence of thermally enhanced cross-relaxation (Figure S8 and S9). The markedly different thermal responses of the $^5D_3$ and $^5D_4$ emissions enable ratiometric temperature sensing based on the $LIR_1$ parameter. The normalized temperature dependence of $LIR_1$ shows that, for low $Tb^{3+}$ concentrations, heating above approximately 320 K results in a rapid, nearly eight-fold increase in $LIR_1$, which stabilizes above 500 K, followed by a slight decrease at higher temperatures (Figure 4e). Increasing the $Tb^{3+}$ concentration progressively suppresses this thermal enhancement of $LIR_1$, as cross-relaxation already significantly affects the emission intensities at low temperatures. This behavior can be quantitatively described by evaluating the relative sensitivity, $S_R$:

$$S_R = \frac{1}{LIR}\frac{\Delta LIR}{\Delta T} \cdot 100\% \qquad (4)$$

where $\Delta LIR$ represents the change of $LIR$ corresponding to the change in temperature by $\Delta T$. Regardless of $Tb^{3+}$ concentration, the temperature dependence of $S_R$ exhibits a single pronounced maximum in the temperature range of 320-420 K (Figure 4f). The maximum sensitivity decreases systematically with increasing dopant concentration, from approximately 1.8% $K^{-1}$ for $LiYO_2$:0.25%$Tb^{3+}$ to about 0.01% $K^{-1}$ for $LiYO_2$:5%$Tb^{3+}$ (Figure 4g). These results demonstrate that variation of the $Tb^{3+}$ concentration allows not only tuning of the emission intensity ratio between the $^5D_3$ and $^5D_4$ levels but also precise control over the maximum $S_R$ of the ratiometric luminescent thermometer based on the $LIR_1$ parameter.

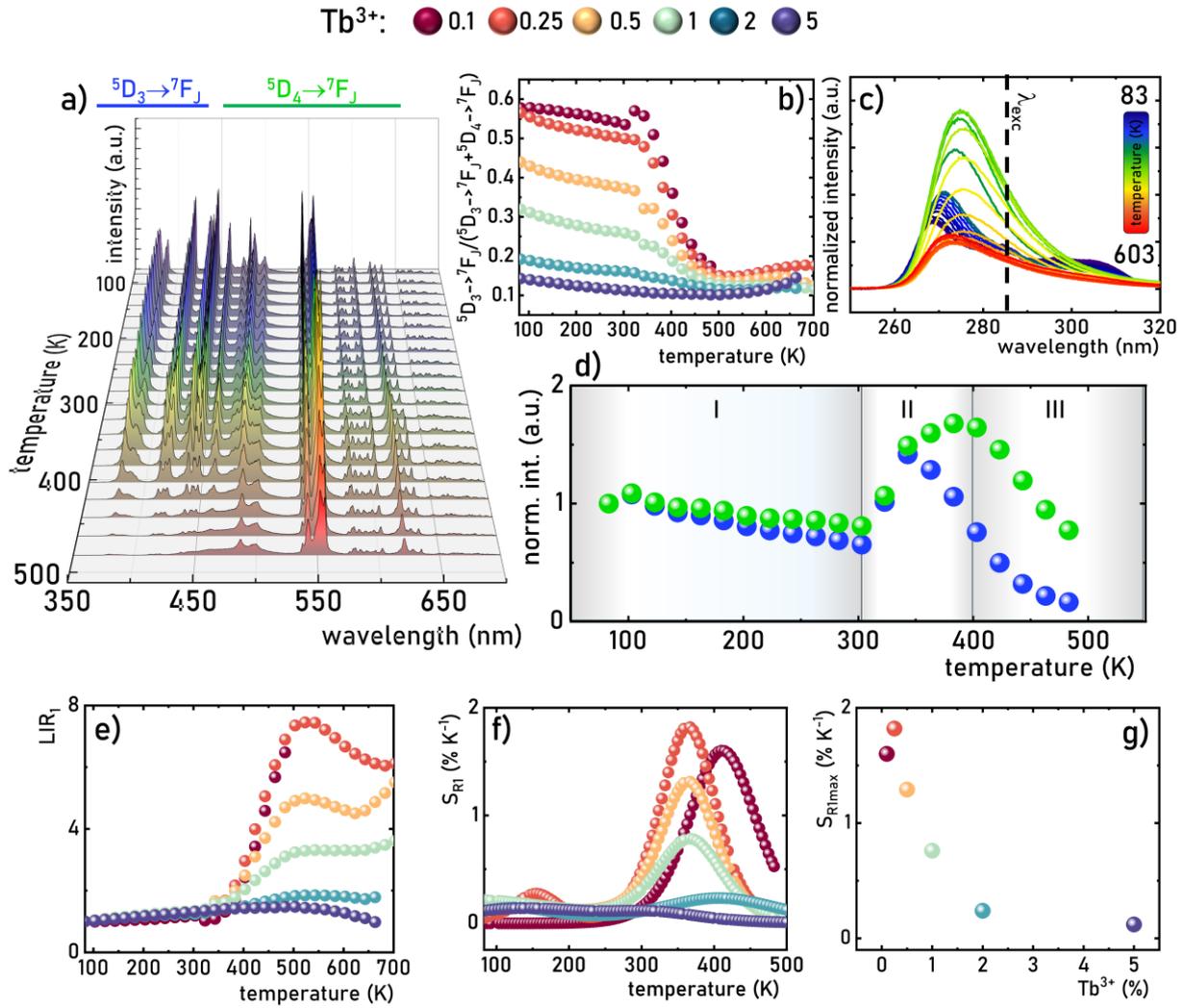

**Figure 4.** Thermal evolution of emission spectra of LiYO$_2$:0.1%Tb$^{3+}$ – a); the thermal dependence of contribution of emission intensity associated with the radiative depopulation of $^5D_3$ state to the total emission intensity of LiYO$_2$:Tb$^{3+}$ for different dopant concentrations – b); thermal evolution of excitation spectra of LiYO$_2$:0.1%Tb$^{3+}$ measured at $\lambda_{em}$ = 543 nm – c); thermal dependence of the emission from $^5D_3$ and $^5D_4$ states for LiYO$_2$:0.1%Tb$^{3+}$ – d); thermal evolution of $LIR_1$ – e) and corresponding $S_R$ – f) for different dopant concentrations; the influence of Tb$^{3+}$ ions concentration on the $S_{R1max}$ – g).

The pronounced temperature-dependent variations in the luminescence intensity of Stark lines associated with the LT and HT phases of LiYO$_2$:Tb$^{3+}$, observed in the vicinity of the phase transition temperature and characterized by high spectral separation, can be effectively exploited for the development of a ratiometric luminescence thermometer. Thermal maps of the normalized luminescence intensity corresponding to emission from the $^5D_3$ level clearly visualize these changes (Figure 5a). To achieve high relative sensitivity in a ratiometric luminescence thermometer, it is advantageous to select two emission signals exhibiting opposite monotonic temperature dependences. Considering that emissions from both the $^5D_3$ and $^5D_4$ levels can be observed in LiYO$_2$:Tb$^{3+}$, the effect how the selection of spectral ranges influences the relative sensitivity of the thermometer was investigated. For emission originating from the $^5D_3$ level, a representative spectral region exhibiting a thermally induced increase in intensity (associated with emission from the HT phase of LiYO$_2$:Tb$^{3+}$) was identified in the 442-444 nm range, whereas a decrease in luminescence intensity (corresponding to the LT phase) was observed in the 440-442 nm range. The ratio of intensities recorded in these spectral windows was used to define the luminescence intensity ratio $LIR_2$:

$$LIR_2 = \frac{\int_{440nm}^{441nm} \left(^5D_3 \to {}^7F_J\right) d\lambda}{\int_{442nm}^{443nm} \left(^5D_3 \to {}^7F_J\right) d\lambda} \qquad (5)$$

With increasing temperature, $LIR_2$ initially exhibits a gradual rise up to approximately 320 K, followed by a rapid increase above this temperature (Figure 5b). The initial increase is most likely attributable to spectral overlap between the LT and HT phase signals as well as temperature-induced spectral broadening. In contrast, the pronounced increase observed near the phase transition temperature is unequivocally related to the structural phase transition of the host lattice. Notably, for higher Tb$^{3+}$ concentrations, the temperature-induced increase in $LIR_2$ near the phase transition becomes significantly less abrupt, which is directly reflected in the corresponding spectral resolution ($S_{R2}$) values (Figure 5c). As expected, the highest $S_{R2}$ values are observed near the phase transition temperature. The maximum $S_{R2}$ value of 10.2% K$^{-1}$ was obtained at 330 K for LiYO$_2$:1%Tb$^{3+}$, whereas for LiYO$_2$:5%Tb$^{3+}$, $S_{R2max}$ decreases to 1.3% K$^{-1}$. These results demonstrate that low Tb$^{3+}$ concentrations enable the realization of one of the few ratiometric phase-transition-based luminescence thermometers operating in the blue spectral region reported to date. Analogous temperature-induced modifications of the emission spectrum associated with the phase transition can also be observed for the $^5D_4 \rightarrow {^7}F_J$ electronic transitions (Figure 5d). Accordingly, an alternative luminescence intensity ratio, $LIR_3$, was defined, yielding a thermal dependence similar to that observed for $LIR_2$ (Figure 5e):

$$LIR_3 = \frac{\int\limits_{492nm}^{494nm} \left(^5D_4 \rightarrow {^7}F_6\right)d\lambda}{\int\limits_{600nm}^{602nm} \left(^5D_4 \rightarrow {^7}F_4\right)d\lambda} \qquad (6)$$

In this case, the corresponding $S_{R3max}$ values are slightly lower, yet remain high. The $S_{R3max}$ value of 8% K$^{-1}$ was obtained for LiYO$_2$:1%Tb$^{3+}$ at 350 K (Figure 5f). As observed previously, increasing Tb$^{3+}$ concentration leads to a reduction in $S_{R3max}$. A detailed analysis of the temperature dependence of the individual signals contributing to $LIR_2$ and $LIR_3$ reveals that the strongest thermally induced decrease in luminescence intensity occurs in the 440-442 nm range, associated with emission from the $^5D_3$ level, whereas the most pronounced thermally induced enhancement is observed in the 600–602 nm range, corresponding to radiative depopulation of the $^5D_4$ level. The differences in the temperature dependence of luminescence signals originating from the LT phase of LiYO$_2$:Tb$^{3+}$ and detected in different spectral regions do not arise from distinct emission mechanisms, but rather from their spectral characteristics. Specifically, when an emission band comprises a larger number of Stark components, related to higher values of the $J$ quantum number, the probability of spectral overlap between Stark lines associated with the LT and HT phases increases. This overlap suppresses temperature-induced signal variations, thereby reducing the dynamic range of the $LIR$ response:

$$LIR_4 = \frac{\int\limits_{442nm}^{443nm} \left(^5D_3 \to {}^7F_J\right) d\lambda}{\int\limits_{600nm}^{602nm} \left(^5D_4 \to {}^7F_4\right) d\lambda} \qquad (7)$$

To maximize the temperature sensitivity of the luminescence intensity ratio to phase-transition-induced changes in LiYO$_2$:Tb$^{3+}$, a new parameter, $LIR_4$, was therefore proposed (Figure 5g). Although the overall temperature dependence of $LIR_4$ resembles that of $LIR_2$ and $LIR_3$, the initial increase in $LIR_4$ is significantly reduced, while a nearly eightfold enhancement is observed in the vicinity of the phase transition temperature. Consequently, the maximum relative sensitivity $S_{R4max}$ reaches 13% K$^{-1}$ for LiYO$_2$:Tb$^{3+}$ (Figure 5h), substantially exceeding the values obtained for $S_{R2}$ and $S_{R3}$. Importantly, with increasing Tb$^{3+}$ concentration, the temperature at which $S_{R4max}$ occurs shifts sublinearly from 317 K for LiYO$_2$:0.1%Tb$^{3+}$ to 355 K for LiYO$_2$:5%Tb$^{3+}$ (Figure 5i). This behavior results from the concentration-dependent modification of the phase transition temperature caused by incorporation of dopant ions with a larger ionic radius than that of the host cation, an effect commonly reported for phase-transition-based luminescence thermometers. Such a correlation was not observed for $LIR_2$ and $LIR_3$, owing to the previously discussed spectral overlap between LT and HT phase emissions, which obscures this trend.

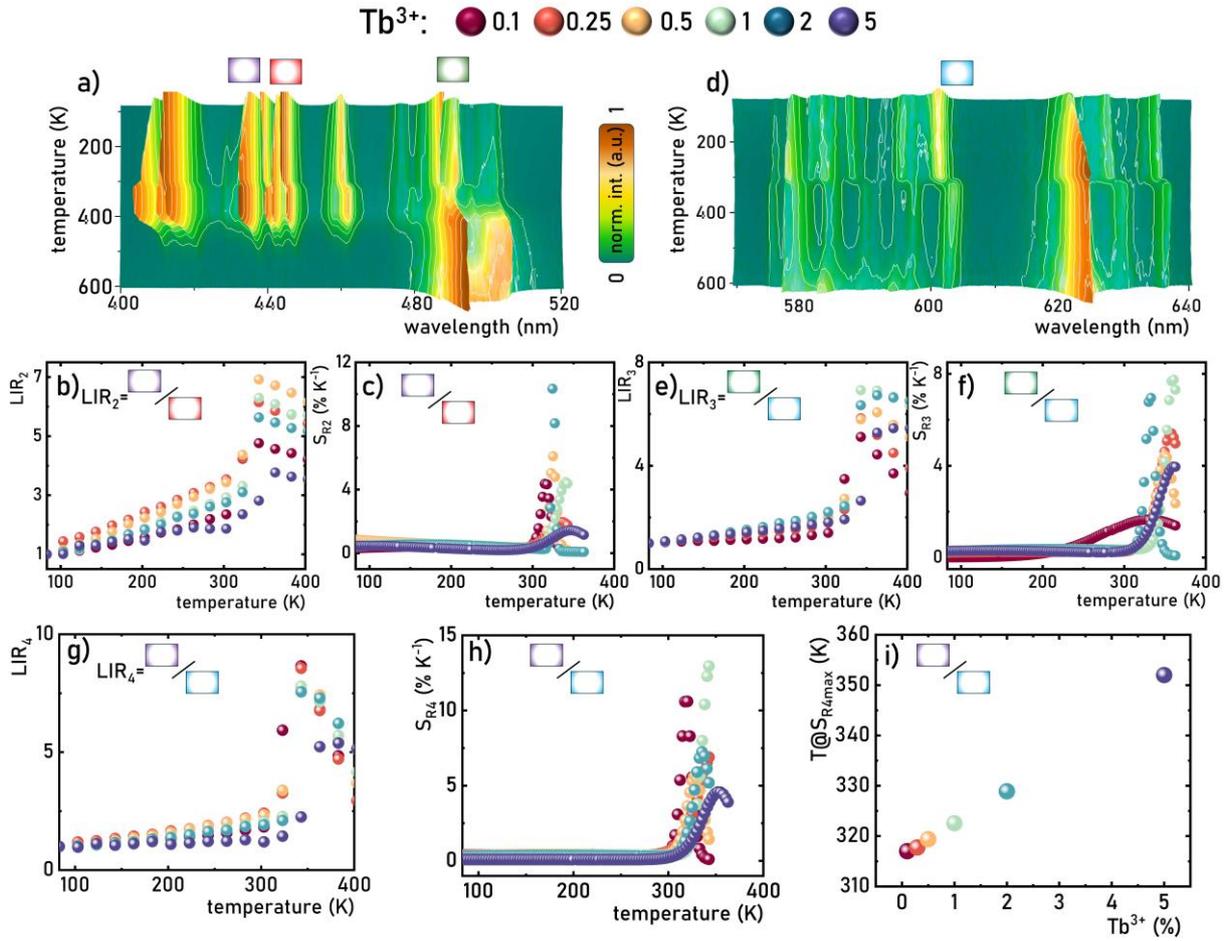

**Figure 5**. Thermal map of normalized emission spectra of LiYO$_2$:0.1%Tb$^{3+}$ in the 400-520 nm– a); thermal dependence of $LIR_2$ – b) and corresponding $S_{R2}$ – c); thermal map of normalized emission spectra of LiYO$_2$:0.1%Tb$^{3+}$ in the 590-640 nm – d); thermal dependence of $LIR_3$ – e) and corresponding $S_{R3}$ – f); thermal evolution of $LIR_4$ – g); and corresponding $S_{R4}$ – h); the influence of dopant concentrations on the $T@S_{R4max}$ – i).

Temperature-induced modifications in the relative contributions of emissions from the $^5D_3$ and $^5D_4$ levels of Tb$^{3+}$ ions in LiYO$_2$:Tb$^{3+}$, together with the fact that the emission bands associated with the $^5D_3 \rightarrow {}^7F_J$ and $^5D_4 \rightarrow {}^7F_J$ electronic transitions occur in distinct spectral regions, enable direct visual detection of temperature variations. The temperature-dependent color evolution of the emitted light for LiYO$_2$:0.1%Tb$^{3+}$ is illustrated in Figure 6a (see also Figures S10 and S11). At 83 K, the emission color is dominated by blue light originating from transitions associated with depopulation of the $^5D_3$ level. With increasing temperature, a pronounced modification of the emission color is observed. Around 343 K, the emission shifts toward a blue-white appearance accompanied by an increase in overall intensity, which is attributed to the temperature-induced modification of the absorption band discussed earlier. At 383 K, the emission color evolves to ivory, and further temperature increases result in a gradual shift toward green emission. These color changes are quantitatively reflected in the evolution of the CIE 1931 chromaticity coordinates (Figure 6b and 6c). For both $x$ and $y$ coordinates, a rapid variation is observed above 300 K, with more pronounced changes occurring in the y coordinate. Increasing the $Tb^{3+}$ ion concentration significantly reduces the dynamics of these temperature-induced chromaticity changes, which can be attributed to the suppression of $^5D_3$ level emission via cross-relaxation processes. Above approximately 450 K, further temperature increases do not result in substantial changes in the chromaticity coordinates, as emission from the $^5D_3$ level is no longer detectable in this temperature range. The dynamics of the chromaticity coordinate variations are directly reflected in the relative sensitivities derived from these parameters. The highest sensitivity values for both coordinates were obtained for $LiYO_2$:0.1%$Tb^{3+}$, with $S_{Rxmax} = 0.40\%$ $K^{-1}$ and $S_{Rymax} = 0.72\%$ $K^{-1}$ at 410 K (Figure 6d and 6e). The dominant role of cross-relaxation processes in governing the concentration dependence of $S_{Rxmax}$ (Figure 6f) and $S_{Rymax}$ (Figure 6g) is further confirmed by the systematic decrease of these sensitivities with increasing $Tb^{3+}$ concentration. For both chromaticity coordinates, higher dopant concentrations lead to a sub-exponential reduction in relative sensitivity.

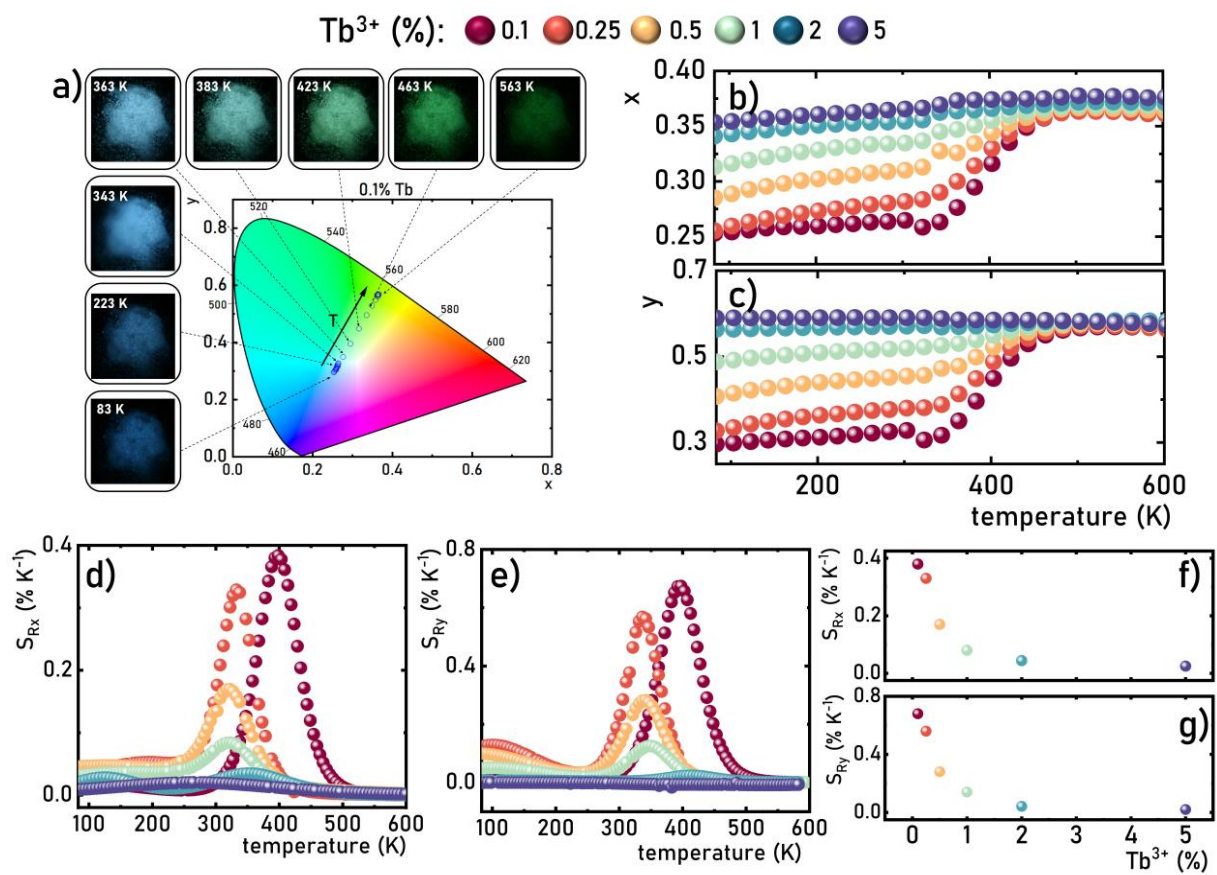

**Figure 6**. The influence of the temperature on the colour of the emission of $LiYO_2:0.1\%Tb^{3+}$ with corresponding CIE1931 chromatic coordinates – a); thermal dependence of chromatic coordinates of x – b) and y – c) for different $Tb^{3+}$ concentration; thermal dependence of $S_{Rx}$ – d); and $S_{Ry}$ – e); the influence of dopant concentration of the $S_{Rxmax}$ – f); and $S_{Rymax}$ – g).

The pronounced temperature sensitivity of $LiYO_2:0.1\%Tb^{3+}$, arising from phase-transition-induced activation of the CR process, highlights the considerable potential of this material for highly sensitive thermal sensing and imaging applications. To experimentally validate this potential, two proof-of-concept demonstrations were performed, illustrating the multimodal functionality of the proposed system. In both experiments, the metallic nozzle of a heat gun was coated with $LiYO_2:Tb^{3+}$ phosphors, and the evolution of their luminescence was monitored under exposure to a controlled flow of heated air (Figure 7a, Figure S12).

In the first experimental configuration, the thermal activation of CR processes was exploited, leading to selective thermal quenching of the $^5D_3 \rightarrow {}^7F_J$ emission. A patterned inscription reading "COLD" was prepared using $LiYO_2$:0.1%$Tb^{3+}$ on a background composed of $LiYO_2$:2%$Tb^{3+}$ (Figure 7a). The choice of the latter as the background material was motivated by the negligible contribution of $^5D_3$-level emission at this dopant concentration (Figures S13 - S20). Under 285 nm excitation, the pattern exhibits intense blue emission, while the background displays green luminescence (Figure 7b). Upon activation of the heat gun, a gradual attenuation of the blue emission is observed. After approximately 32 s of heating, the inscription becomes indistinguishable, and the entire surface emits uniformly in the green spectral region. This process is fully reversible: upon switching off the heat source (and turning on the cooling mode) after approximately 40 s, the blue emission progressively reappears and becomes clearly visible again within 58 s (Figure S21). A complementary control experiment using a thermovision camera confirms that the disappearance of the blue emission coincides with the local temperature exceeding approximately 330 K (Figure S22). These observations demonstrate that thermally activated CR in $LiYO_2$:$Tb^{3+}$ can be exploited to construct intuitive luminescent safety indicators, capable of providing unambiguous visual information about critical temperature thresholds, thereby reducing the risk of accidental burns and enabling safe human-machine interaction. In this context, $LiYO_2$:$Tb^{3+}$ functions effectively as a thermally triggered luminescent switch.

The full capabilities of phase-transition-induced CR activation in $LiYO_2$:$Tb^{3+}$ are further revealed in dynamic thermal imaging experiments. Luminescence images of $LiYO_2$:0.1%$Tb^{3+}$ acquired as a function of temperature exhibit pronounced spectral redistribution: emission associated with the $^5D_3 \rightarrow {}^7F_J$ transitions is detected exclusively in the blue (B) channel of a standard digital camera, whereas emission from the $^5D_4$ level is predominantly recorded in the green (G) channel (Figure 7c). This intrinsic separation of spectroscopic information enables filter-free, two-dimensional thermal imaging using conventional digital camera. For this demonstration, a narrow strip of $LiYO_2$:0.1%$Tb^{3+}$ phosphor (5.5 cm in length) was deposited along the airflow direction on the metal nozzle of the heat gun (Figure S23-S26). Following activation of the heat source, luminescence images were captured at 15 s intervals. From each image, the intensity distributions in the B (Figure S24) and G (Figure S25) channels were extracted and combined to generate two-dimensional G/B ratio maps (Figure 7d). Using an experimentally established calibration curve (Figure 7e, see equation S1), these ratio maps were directly converted into spatially resolved temperature distributions, allowing real-time thermal mapping of the heated component using only a standard camera (Figure 7f). Analysis of the resulting thermal maps indicates that the maximum temperature, located near the air outlet of the nozzle, increases to approximately 510 K after 60 s of heating (Figure 7g). Upon turning off of the heat gun, a gradual cooling process is observed, proceeding at a significantly slower rate than the heating stage. The temperature values and spatial gradients derived from the luminescence-based method show excellent agreement with those obtained independently using a commercial thermovision camera (Figure S27). Cross-sectional analysis along the phosphor strip further confirms that, at early heating stages (t < 30 s), the highest temperatures are localized near the outlet, in full accordance with thermovision data (Figure 7h).

Collectively, these results clearly demonstrate that phase-transition-induced activation of cross-relaxation in $LiYO_2$:$Tb^{3+}$ can be effectively harnessed for the construction of simple, visually intuitive safety indicators as well as for high-contrast, filter-free thermal imaging. Importantly, the temperature at which CR becomes activated, directly linked to the phase transition temperature of the host lattice, can be tuned over a broad range, as shown in previous studies[16,20]. This tunability enables precise adjustment of the operating temperature window, allowing optimization of the luminescent thermal marker for application-specific requirements.

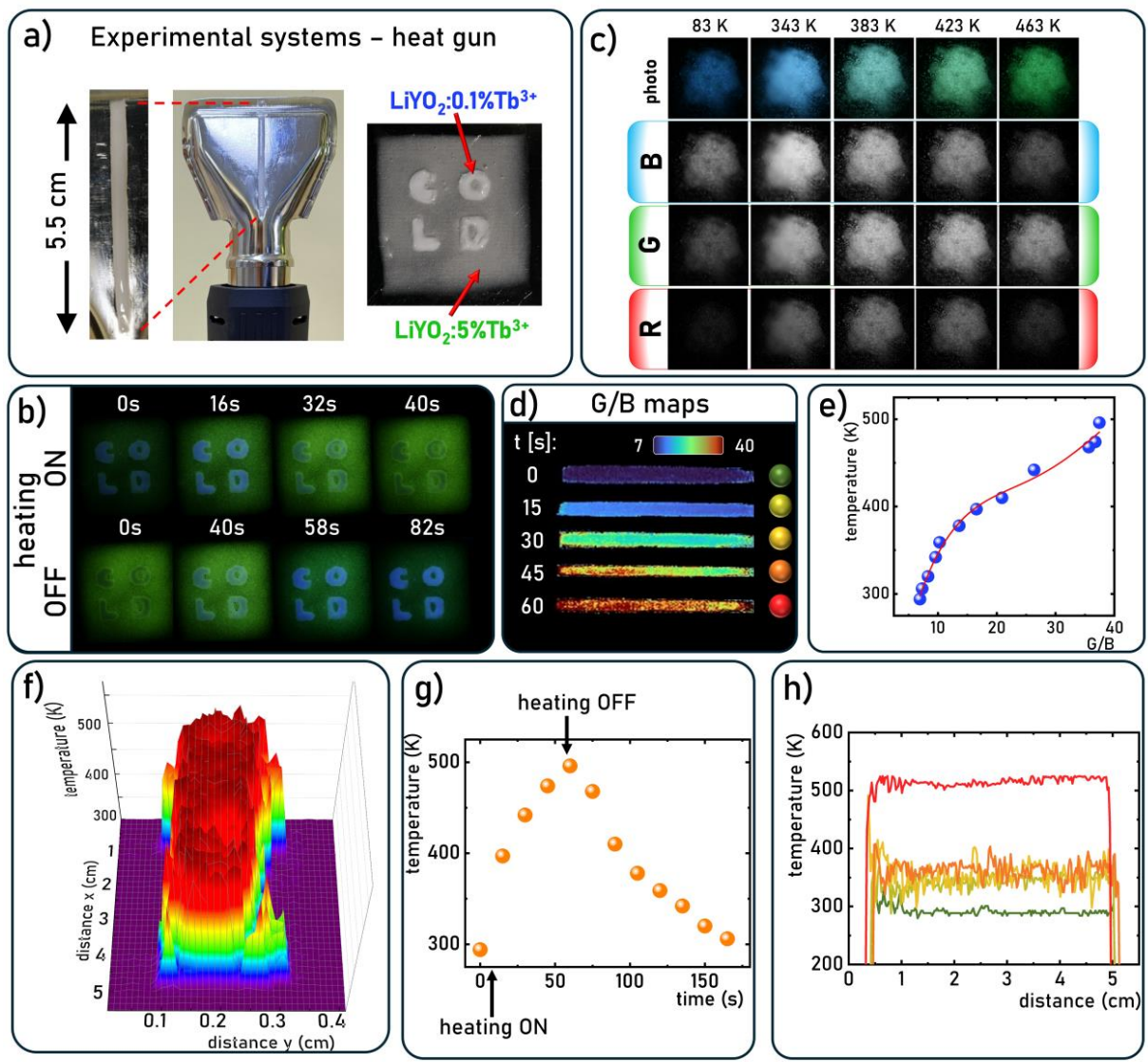

**Figure 7**. The experimental setup used in the thermal imaging experiment – a); photos of the luminescence of the pattern deposited on the metal element of heat gun as a function of time after turning on and turning off heating – b); the representative photos of luminescence of the LiYO$_2$:0.1%Tb$^{3+}$ at different temperatures and corresponding B, G, R maps – c); the G/B maps of the strip of the LiYO$_2$:0.1%Tb$^{3+}$ deposited at the metal element of the heat gun obtained at different times after turning on the heating– d); calibration curve of the temperature as a function of G/B ratio– e); 3D map of temperature distribution of on the LiYO$_2$:0.1%Tb$^{3+}$ strip after 60 s of heating of heat gun– f); the average temperature on the metal element of a heat gun determined

based on the G/B luminescence thermometry as a function of time – g); and temperature distribution in the cross section of the LiYO$_2$:0.1%Tb$^{3+}$ strip at different times of heating (colours of the curves corresponds to the assignments in Figure 7d).

**Conclusions**

This work presents a systematic investigation of the influence of temperature and Tb$^{3+}$ ion concentration on the spectroscopic properties of LiYO$_2$:Tb$^{3+}$, with the objective of developing a multimodal visual luminescent thermometer with high relative sensitivity. LiYO$_2$ undergoes a structural phase transition from a low-temperature monoclinic phase to a high-temperature tetragonal phase at temperatures above approximately 320 K. This transition has a pronounced impact on the spectroscopic behavior of Tb$^{3+}$ ions incorporated into the host lattice. For low dopant concentrations, exemplified by LiYO$_2$:0.1%Tb$^{3+}$, the emission spectrum is dominated by intense bands originating from radiative transitions from the $^5$D$_4$ excited state to the $^7$F$_J$ manifold, accompanied by weaker emission associated with transitions from the $^5$D$_3$ level. Increasing the Tb$^{3+}$ ion concentration results in a gradual reduction of the emission intensity from the $^5$D$_3$ state relative to that from the $^5$D$_4$ state. This behavior arises from an enhanced probability of cross-relaxation between neighboring Tb$^{3+}$ ions, driven by a reduction in the average Tb$^{3+}$-Tb$^{3+}$ distance. The cross-relaxation process is evidenced by the near invariance of the average lifetime of the $^5$D$_4$ level with increasing dopant concentration, together with a pronounced shortening of the average lifetime of the $^5$D$_3$ level. Investigation of the temperature dependence of the $^5$D$_3$ emission intensity reveals that it remains nearly temperature independent up to approximately 300 K. Above this temperature, a moderate increase in emission intensity is observed, followed by rapid thermal quenching. The onset temperature for this quenching coincides closely with the structural phase transition temperature of LiYO$_2$. The increase in lattice symmetry accompanying the phase transition reduces the number of Stark components within the $Tb^{3+}$ multiplets, while simultaneously increasing their energetic separation due to a stronger crystal field. These changes facilitate the energetic conditions required for cross-relaxation to occur. The thermal enhancement of emission intensities from both the $^5D_3$ and $^5D_4$ levels in the vicinity of the phase transition temperature is attributed to phase-transition-induced modifications of the excitation spectra of $Tb^{3+}$ ions. Differences in the thermal evolution of emission from the $^5D_4$ and $^5D_3$ levels enable ratiometric luminescence thermometry, yielding a maximum relative sensitivity of 1.8% $K^{-1}$ for $LiYO_2$:0.25%$Tb^{3+}$ at 320 K. Furthermore, temperature-dependent changes in the shape of the emission spectrum associated with the structural phase transition, observed for both the $^5D_3 \rightarrow {}^7F_J$ and $^5D_4 \rightarrow {}^7F_J$ transitions, enabled the development of three distinct temperature readout modes. Maximum relative sensitivities of 10.2% $K^{-1}$ at 330 K for $LiYO_2$:1%$Tb^{3+}$ operating in the blue spectral range and 8% $K^{-1}$ at 350 K in the green spectral range were achieved. When emission signals from both excited states are simultaneously considered, sensitivities as high as 13% $K^{-1}$ can be obtained. Thermally activated cross-relaxation also induces a continuous color shift of the emitted light from blue through blue-white to green. This temperature-dependent shift in chromaticity coordinates enables visual temperature readout, yielding the highest chromatic sensitivity reported to date for $LiYO_2$:0.1%$Tb^{3+}$, with $S_{Rxmax}$=0.40% $K^{-1}$ for the x coordinate and $S_{Rymax}$=0.72% $K^{-1}$ for the y coordinate at 410 K. This effect was exploited to demonstrate two thermal readout modes. In the first mode, temperature-dependent changes in the intensities recorded in the blue and green channels of a digital camera enabled filter-free, two-dimensional thermal imaging of temperature gradients across a metallic element during heating and cooling. In the second mode, deliberate spatial distribution of $LiYO_2$:$Tb^{3+}$ phosphors with different $Tb^{3+}$ concentrations enabled the design of simple and highly intuitive temperature-exceedance indicators for safety applications.

The results presented in this study demonstrate, for the first time, that a structural phase transition can activate cross-relaxation processes, thereby rapidly altering the thermometric performance of a luminescent thermometer. This phenomenon enables the realization of a multimodal luminescent thermometer with a broad range of potential applications.

**Acknowledgements:**

This work was supported by the National Science Center (NCN) Poland under project no. DEC-UMO-2022/45/B/ST5/01629. M. Sz. gratefully acknowledges the support of the Foundation for Polish Science through the START program. Authors would like to acknowledge dr Justyna Zeler for help in samples annealing and dr Damian Szymanski for SEM analysis.